\documentclass[lettersize,journal]{IEEEtran}
\usepackage{cite}
\usepackage{amsmath,amssymb,amsfonts}
\usepackage[linesnumbered,ruled,vlined]{algorithm2e}
\usepackage{graphicx}
\usepackage{textcomp}
\usepackage{xcolor}
\usepackage{tabularx}
\usepackage{subcaption}
\usepackage{multirow}
\usepackage{tabularx}
\usepackage{url}
\usepackage{stfloats}
\usepackage{array}
\usepackage{comment}
\def\BibTeX{{\rm B\kern-.05em{\sc i\kern-.025em b}\kern-.08em
    T\kern-.1667em\lower.7ex\hbox{E}\kern-.125emX}}
\begin{document}

\title{Blockchain-Enabled Device-Enhanced Multi-Access Edge Computing in Open Adversarial Environments\\
}

\author{\IEEEauthorblockN{Muhammad Islam,}
\and
\IEEEauthorblockN{Niroshinie Fernando,}
\and
\IEEEauthorblockN{Seng W. Loke,}
\and
\IEEEauthorblockN{Azadeh Ghari Neiat,}
\and
\IEEEauthorblockN{Pubudu N. Pathirana} \\
\IEEEauthorblockA{School of Information Technology, Deakin University, Geelong, Australia}
}

\maketitle

\begin{abstract}

We propose Blockchain-enabled Device-enhanced Multi-access Edge Computing (BdMEC). BdMEC extends the Honeybee framework for on-demand resource pooling with blockchain technology to ensure trust, security, and accountability among devices (even when they are owned by different parties). BdMEC mitigates risks from malicious devices by making computations traceable. Our prototype and results demonstrate BdMEC's ability to manage distributed computing tasks efficiently and securely across multiple devices. 


\end{abstract}


\section{Introduction}

\begin{figure}[tp]
\centerline{\includegraphics[width=\linewidth]{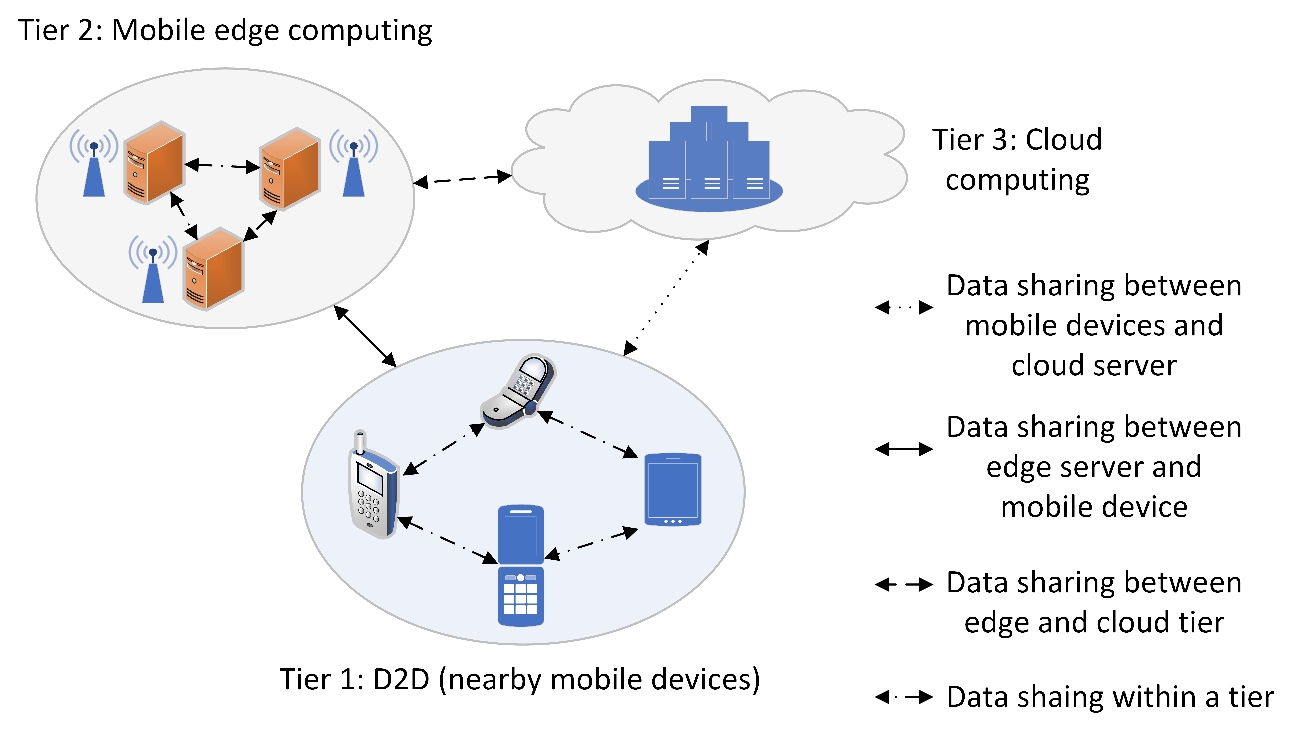}}
\caption{Different tiers involved in device-enhanced MEC.}
\label{layers}
\end{figure}

Complementing conventional edge computing, mobile devices can leverage the idle resources of other nearby mobile devices  \cite{honeybee}. Those devices are usually in closer proximity to each other compared to the cloud or even edge servers, as illustrated in Fig. \ref{layers}. This paradigm is called `device-enhanced multi-access edge computing' (or device-enhanced MEC)~\cite{fang2021joint} and aims to supplement the cloud and edge computing servers, by reducing their burden while improving overall performance. 

Depending on the connectivity and resource availability, recent research has shown that device-enhanced MEC can provide faster processing of resource-intensive tasks and energy avings~\cite{fang2021joint,nagesh2022opportunistic}. 
However, the device-enhanced MEC paradigm also presents challenges related to integrity, security, and privacy, that remain largely unresolved in the literature to the best of our knowledge. Sharing application data and tasks with other devices (owned by different users)  introduces risks to the integrity of the application, since it becomes challenging to ascertain the honest behaviour of all participating nodes as there is a possibility of malicious nodes compromising the integrity of the application \cite{comstsurvey}.
Mobile devices can engage in various malicious activities, including  data tampering, unauthorized access, denial of service (DoS) attacks, malware distribution and resource exhaustion.  
Therefore, it is essential to implement security measures and protocols to detect and mitigate such activities to ensure the integrity and overall security of the system. 

Blockchain is a rapidly evolving technology that enables decentralized and transparent data storage by securely recording each activity in a cryptographically linked and immutable manner, ensuring the integrity and immutability of the recorded data \cite{b6}. Existing works have explored similar approaches for utilizing idle resources of nearby mobile devices, as discussed in works such as \cite{multipersona,d2dassisted}. Among these, the most relevant approach has been presented in \cite{d2dassisted}, which proposes a blockchain-based resource trading and sharing scheme to leverage the idle resources of nearby mobile devices. 
Yet, existing methodologies have overlooked several critical concerns:
\begin{enumerate}
\item \textbf{Malicious Activities:}  While \cite{d2dassisted} tackles the trust issues between mobile devices, it does not address the problem of malicious activities such as intentionally slowing down the task execution by a nearby mobile device and tracking of devices by the greedy participants of the network.
\item \textbf{Overhead of Resource Registration and Requesting:} In \cite{d2dassisted}, mobile devices are required to register their resources on a smart contract which are then requested by other mobile devices in need of resources. This registration and request process adds extra overhead to the system, potentially affecting efficiency.
\item \textbf{Dependency on Stable Network:} A stable mobile edge computing network is necessary for mobile devices as discussed in \cite{d2dassisted} to register their resources on the system effectively; otherwise, those resources cannot be utilized. However, frequent changes in nearby mobile devices' locations make it impractical to maintain a stable network. 
\end{enumerate}
These aforementioned issues hinder the practicality and effectiveness of existing approaches in device-enhanced MEC contexts, calling  for an enhanced mechanism that addresses these challenges and leverages the benefits of blockchain technology to enable secure, efficient, and reliable resource utilization in such device-enhanced MEC settings.  

In this regard, we propose a blockchain-enabled device-enhanced MEC mechanism (BdMEC). In BdMEC, we perform two distinct types of operations: (1) the front end focuses on the connectivity of nearby mobile devices, and (2) the backend handles the recording of application's offloading related statistics on the blockchain network such as the type of application, the number of connected nearby mobile devices, and the resource contribution made by each device in shared computation. To quantify the effectiveness of resource utilization, we use speed gain, denoted as $\bold{S} = \frac{time_1}{time_2}$, where $time_1$ is the time taken by a single mobile device to complete the application, and $time_2$ is the time taken to complete the application while utilizing idle resources of nearby mobile devices. Ideally, we aim for a high value of speed gain $\bold{S} >1$  which shows that idle resource utilization has reduced the completion time of the given application. 
To identify malicious mobile nodes, we use a calculated fractional speed gain $\bold{S}_i$ for each \(i^{\text{th}}\) device, via a weighted average of historical speed gain (\(S\)) values from each task participation, incorporating task complexity and the number of jobs as weights. The resulting $\bold{S}_i$ offers an estimate of a device's likelihood to enhance task completion speed in future collaborations.
%
%
Hence, if $\bold{S}_i<1$ then the device is regarded as malicious. It is important to note that this method also categorizes slow or weak mobile devices as `malicious', as they would contribute to a low-speed gain. In addition, we propose a privacy model based on differential privacy \cite{dpint2006} for BdMEC to avoid further disruption from malicious and greedy devices. 
In particular, we make the following contributions: 
\begin{itemize}
\item We propose a blockchain-enabled device-enhanced MEC mechanism (BdMEC) that integrates blockchain into an existing device-enhanced MEC work-sharing model~\cite{honeybee} to leverage the idle resources of nearby mobile devices. BdMEC incorporates a querying system to check the historical task statistics recorded on the blockchain, thereby avoiding nodes with a history of malicious behavior before initiating work sharing or application-related data. 
\item We propose a new privacy model for BdMEC using differential privacy to avoid malicious devices and greedy behaviour. A controlled noise is added to the number of jobs executed by mobile devices before writing it to the blockchain network, thereby fooling the greedy devices.  
\item We validate the effectiveness of the proposed BdMEC experimentally using actual device-enhanced MEC testbed data conducted in a network scenario using Android smartphones. The results demonstrate that BdMEC achieves a speed gain $\bold{S} > 1$ even under network connectivity with high uncertainty and mobility, showing that BdMEC is a viable approach even considering the added overheads of Blockchain integration. 
\end{itemize}

Subsequently, Section II introduces   the related concepts and  the proposed Blockchain-enabled device-enhanced MEC mechanism (BdMEC).  Section III presents the simulation setup and evaluation results. Section IV concludes the paper and highlights future works.

\section{Our Approach}
\subsection{Honeybee Model}
The proposed architecture employs the Honeybee model and framework ~\cite{honeybee} to support the device-enhanced MEC paradigm. This model harnesses the collective power of mobile devices as edge computing resources, allowing them to share their workload and improve performance while conserving energy. In the Honeybee model, the device responsible for distributing tasks is referred to as the ‘delegator'. The delegator first divides the task into individual ‘jobs', and simultaneously initiates the resource discovery process to identify suitable nearby ‘worker' mobile devices for offloading those jobs. These jobs are stored in a shared job queue that is accessed synchronously by the delegator and participating workers. The nodes are required to proactively ‘steal' jobs from the shared job queue. Honeybee uses a work stealing algorithm to automatically balance the jobs among the delegator and the workers based on factors such as job completion rates and transmission delays, allowing more capable nodes to ‘steal' more jobs. This method has proven effective in achieving performance gains and energy savings even in situations where nodes join or leave dynamically.

The Honeybee framework\footnote{https://github.com/niroshini/honeybee} is an Android implementation of the Honeybee model. In this work, we have chosen to use the Honeybee framework to investigate the feasibility of integrating blockchain into a device-enhanced MEC paradigm. This decision was motivated by these factors: a) Honeybee has experimentally been shown to give performance gains and energy savings across various conditions~\cite{honeybee}; b) it is open-source;  c) it includes mechanisms for automatic load-balancing and fault-tolerance, enhancing system robustness; d) it supports ad-hoc connections, eliminating the need for nodes to be known/established a prior; and e) Honeybee has been shown to be extensible, as evidenced by extensions for drones, robots, and dependency-based task scheduling~\cite{nagesh2022opportunistic}.

\begin{figure}[tp]
\centerline{\includegraphics[width=\linewidth]{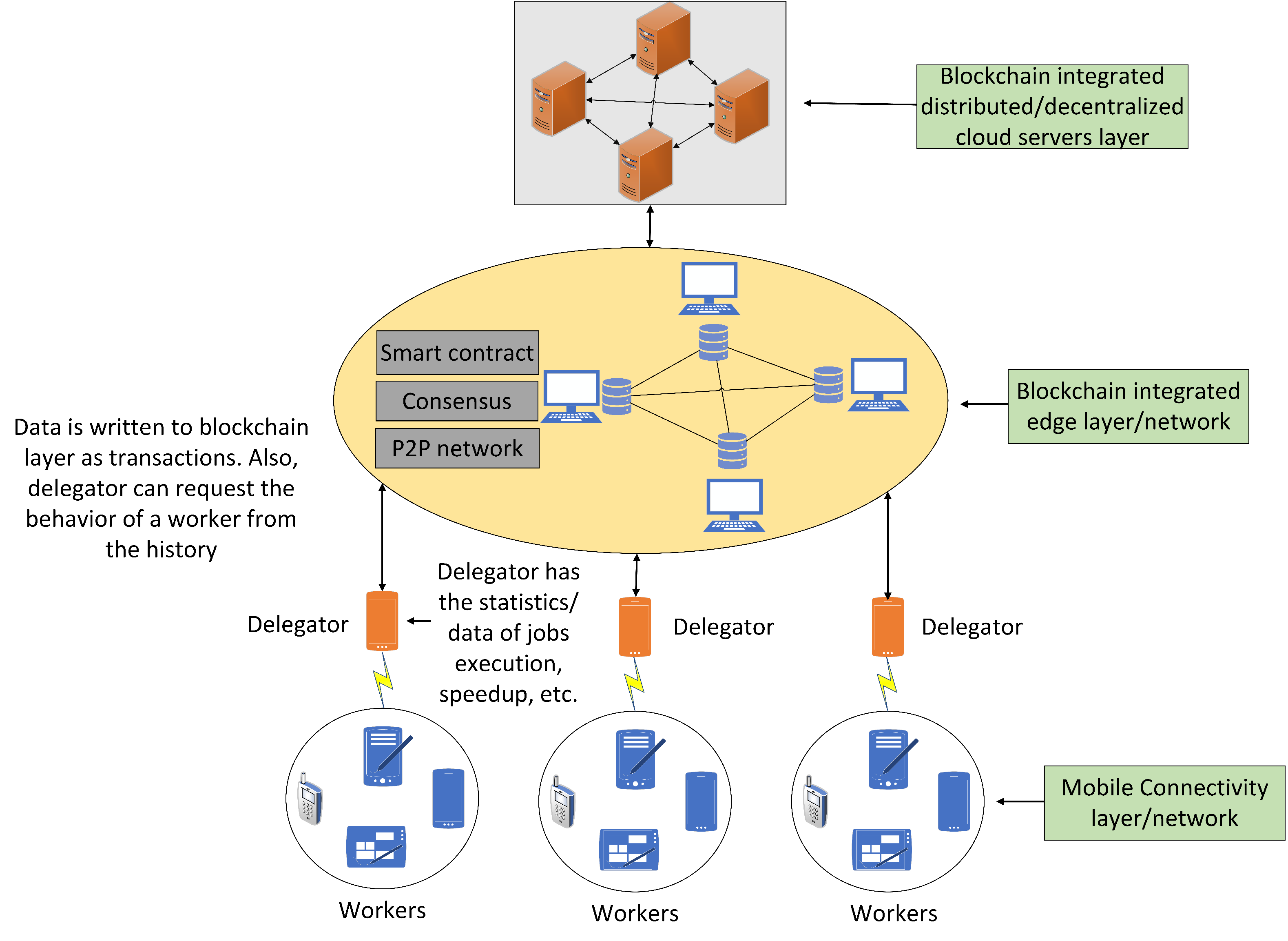}}
\caption{Architecture of the proposed blockchain-enabled device-enhanced MEC mechanism (BdMEC).}
\label{system_model}
\end{figure}

\subsection{Proposed Architecture}
In this section, the proposed architecture for blockchain-enabled device-enhanced MEC (BdMEC) is discussed in detail. The architecture is presented in Fig. \ref{system_model} which consists of the following components. 
\paragraph{Task} A task is a resource-intensive application such as searching the number of matches for a picture of a particular person in a set of 1000 or even more different images. Similarly, a task can also be any other sort of computation-intensive application which usually will take more than one minute to be completed by the initiating mobile device by itself. Furthermore, the task can be divided into chunks such as a chunk of 40 pictures or images, and each image is called a job. The time taken by a single task execution is regarded as iteration.
\paragraph{Delegator node} A delegator node is the initiating mobile device, where the task originates from, and seeks extra computation resources to efficiently execute resource-intensive tasks or heavy applications. The delegator has the ability to search for other mobile devices within its D2D connectivity range (such as via Bluetooth or Wi-Fi-Direct). Furthermore, a delegator can connect with many worker nodes under the limit of the particular D2D protocol used. In the experiments in this paper, Wi-Fi Direct is used in D2D communications.
\paragraph{Worker node} A worker node is a nearby mobile device in the vicinity of the delegator node that is idle and thus can help the delegator finish the resource-intensive task quickly. Moreover, a worker node also has the ability to get connected with a delegator node when requested. 
\paragraph{Mobile connectivity layer} This layer deals with the D2D wireless connectivity between delegator and worker nodes. Furthermore, the delegator uses Wi-Fi Direct to connect with the worker node and share a portion of a task, i.e., a job or number of jobs.  

\paragraph{Blockchain layer} This layer comprises edge servers accessible by delegator nodes. Each edge server functions as a blockchain node, recording task statistics such as executed jobs and worker performance. The distributed nature of edge servers ensures accessibility from various regions via Wi-Fi or cellular connectivity. Task statistics are recorded as transactions in the blockchain for reliable tracking and auditing of job distribution and worker performance. The edge server responds to delegator queries, providing information about a specific worker to predict its behavior. 
\paragraph{Cloud layer} The cloud layer serves as the backend storage mechanism in the system architecture. It is responsible for handling the increasing volume of transactions and data generated by the edge servers. As the number of transactions grows, the edge servers offload excess data to the cloud, effectively mitigating storage limitations that could arise at the blockchain layer.

\subsection{Privacy Model}
\label{sec:priv}
In BdMEC, the privacy of a worker is defined as follows. 

\textbf{Definition 1:} \textit{The job count recorded on a worker's blockchain history, which accumulates over time, plays a crucial role in their selection for future iterations. This information is sensitive and should be safeguarded to maintain the worker's privacy and integrity within the network.}

The privacy model employs two separate ledgers: the delegator's ledger and the worker's ledger. To achieve this, we utilize the Hyperledger Fabric multi-ledger capability \cite{hyperledger}, which enables nodes in the blockchain network to be organized into distinct channels, each with its own private ledger. Nodes within the same channel can only access their respective ledger and not the ledgers of other channels. The delegator's ledger, which is only accessible to delegators, is managed by channel1 and is used for tracking the job count of workers. Conversely, the worker's ledger, which is public and accessible to both delegators and workers, is used for recording the noisy count of workers. This ensures that adversarial nodes cannot track the actual records of honest nodes and predict the next selected worker accurately.

Differential privacy, a widely used method for safeguarding sensitive information in statistical databases, is employed in our approach. We use the Laplace mechanism for calibrated noise addition~\cite{dpint2006}. To evaluate the noise magnitude we will use relative error \textbf{R} which is defined as follows.

\textbf{Definition 2:} \textit{Relative error \textbf{R} is the ratio of absolute difference between actual value $v$ and perturbed/noise-added value $v^{'}$. Mathematically it is defined as follows.}

\begin{align}
\textbf{R} = \frac{|v - v^{'}|}{v} \times 100\%
\label{eq:er}
\end{align}

Consequently, the introduction of random noise results in noisy job counts.
To evaluate the impact of noise addition, we use precision \textbf{P} which is defined as follows:
\begin{align}
\textbf{P} = \frac{\text{true positive}}{\text{true positive + false negative}} \times 100\%
\label{eq:prec}
\end{align}
Here, true positive refers to a worker holding the highest job count in all records, which aligns with the top position on both the delegator's and worker's ledgers. False negative indicates a worker with a significant job count on the delegator's ledger, but it is placed lower due to random noise. Precision is a percentage that gauges the system's accuracy in recognizing true positives. A precision of 100\% indicates perfect identification and recording of high job counts in both ledgers without any errors.

\subsection{Adversarial model}
\label{sec:adv}
In a typical device-enhanced MEC context, two types of adversaries can be considered; Internal and External adversaries.
\textit{Internal adversaries} can be worker devices within the Honeybee network. For instance, malicious workers can send incorrect results, such as a different number of matches for a particular picture, or delay sharing the result, which significantly impacts the speed gain. Additionally, malicious and greedy workers can track honest workers and manipulate records on the blockchain, such as predicting which workers will be selected in the next round. The predicted workers can be targeted with attacks such as Sybil or 51\% attack \cite{b6} to isolate their contribution in the consensus. As a result, we need a mechanism to track and avoid workers' malicious behavior.
\textit{External adversaries} are devices that have a curious nature, seeking unauthorized access to the Honeybee network, aiming to disrupt its operation. Their actions may involve causing disconnections among connected devices, unauthenticated access to photos/images, and stealing other application-related data such as pictures or other sensitive information. Thus protective measures against such malicious activities perpetrated by external adversaries are  crucial for the smooth operation of the device-enhanced MEC network. 
%

Furthermore, based on the risk level, the adversaries can be classified as (1) honest, (2) dishonest-but-not-curious, and (3) dishonest-and-curious, as detailed below. 
\begin{enumerate}
 \item Honest: internal nodes that are 100\% honest and not engaging in any of the aforementioned malicious activities.
 \item Dishonest-but-not-curious: internal nodes that will not steal any information. However, they can cheat by sending an incorrect number of processed tasks to the delegator. For instance, dishonest-but-not-curious nodes can send a greater number of images than processed to delegators to get more financial benefit or they can send incorrect ids of the jobs executed.   
 \item Dishonest-and-curious: external nodes that intend to steal data/information, get unauthenticated access to the Honeybee network to perform malicious activities, and disrupt the operation of the network.
\end{enumerate}
In this work, we focus on mitigating the potential malicious activities of dishonest-but-not-curious nodes (internal adversaries) within the Honeybee network. 
Future research will focus on developing strategies to protect against the malicious actions of dishonest-and-curious nodes (external adversaries).

\begin{algorithm}[b]
\small
\let\oldnl\nl
\newcommand{\nonl}{\renewcommand{\nl}{\let\nl\oldnl}}
\caption{Selection of a worker with high capacity and honest behavior}
\label{alg:alg1}
\SetAlgoLined
\DontPrintSemicolon
\nonl
\textbf{Input:} History recorded on blockchain, worker behavior $\lambda$, workers pool\;
\nonl
\textbf{Output:} Decision on whether the delegator is executing the tasks locally or selecting workers to share the execution, which workers from the workers' pool are selected for current tasks/application, value of $\lambda$\;
\nonl
\textbf{Initialization:} Index $i = 0$, ID of worker, no. of jobs executed, steal chunk size, speed gain of worker $i$ $\bold{S_{i}}$, workers pool $\omega = \{W_{1}, W_{2}, W_{3}, ..., W_{n}\}$, $\lambda_{i} = \pm1$\;
\While {\textit{Invoke}} {
    $\omega_{sorted} \leftarrow$ \textbf{Call} FindMax()\;
    $\bold{S_{i}} \in \omega_{sorted}$\;
    \If {$\bold{S_{i}} > 1$ and $\lambda_{i} > 0$} { 
        $W_{i}$ is selected from the pool $\omega$\;
        \KwRet\;
    }
    \Else{
             The delegator proceeds with local execution\; 
             \KwRet\;
    }

    \nonl
   $\textbf{FUNCTION}\rightarrow FindMax()$\;  
   \Indp
    \textit{sort $\omega$ based on values of speed gain}\;
    \KwRet {$\omega_{sorted}$}\;
    \Indm
} 
\end{algorithm}

\subsection{Working of the Proposed blockchain-enabled device-enhanced MEC Mechanism (BdMEC)}
\label{sec:working}
BdMEC operates by utilizing a pool of workers, denoted as $\omega = \{W_{1}, W_{2}, W_{3}, ..., W_{n}\}$, where n represents the maximum number of workers in the pool. The aim of the delegator is to achieve a speed gain $\bold{S} > 1$ for faster task execution. To achieve this, the delegator must distinguish between high and low capacity workers. Workers with high capacity result in $\bold{S} > 1$, whereas those with low capacity degrade performance. Additionally, BdMEC maintains a variable $\lambda$ on the blockchain for each worker, which evaluates the risk of a worker being malicious. The variable has two possible values of +1 and -1, referred to as the behavior factor. A value of $\lambda_{i} = +1$ indicates no malicious activity, while $\lambda_{i} = -1$ indicates potential malicious behavior. Such behaviors include claiming inaccurate job completion numbers, fabricating job details such as job IDs, repeated failure to complete assigned tasks, and a history of malicious reports within the network. If at least one malicious behavior is noted, then $\lambda_{i}$ is set to -1. Conversely, if no malicious behavior is observed, then the delegator sets $\lambda_{i} = 1$. Before sharing computation resources, the delegator sends a query to the blockchain network to obtain the value of $\lambda_{i}$ for worker $W_{i}$ in the pool. The smart contract in BdMEC evaluates the query over the record in the blockchain.

Algorithm \ref{alg:alg1} presents the mathematical steps for selecting workers with high capacities. The process involves calculating each worker's `fractional contribution', denoted as $Wc_{i}$, by examining their historical performance in previous offloading iterations. Subsequently, a `fractional speedup', $\bold{S_{i}}$, is calculated for each worker using the $Wc_{i}$, serving as a metric to identify the best workers for task offloading. $\bold{S_{i}} > 1$ indicates that $W_{i}$ has high capacity, while $\bold{S_{i}} \leq 1$ indicates that $W_{i}$ has low capacity and should not be selected for the current iteration. In Line 1 of the algorithm \ref{alg:alg1}, the smart contract is invoked. In Lines 2-14 of algorithm \ref{alg:alg1}, workers with high capacities are evaluated, and the value of $\lambda_{i}$ is used to check the behavior of worker $W_{i}$. Consequently, the worker is selected from the pool if $\bold{S_{i}} > 1$ and $\lambda_{i} > 0$. If the aforementioned condition is not satisfied, the worker is not selected, and the delegator continues execution of the current task or application on their own. Thus, algorithm \ref{alg:alg1} enables the delegator to select workers with high capacity and honest behavior.

After successfully completing a task or application, the parameters, including the number of jobs executed, speed gain, location, and steal chunk size, along with the IDs of worker devices, are recorded in the delegator's blockchain ledger. This allows the delegator to select workers with high capacities and honest behavior. However, to prevent greedy workers from seeing the actual number of jobs executed, the count is perturbed and controlled noise is added to it before it is recorded in the worker's ledger. If the delegator is not connected to the blockchain network during task execution, the statistics are held until the connection becomes available. Therefore, writing data to a blockchain network does not significantly impact speed gain. In the following section, we will evaluate the proposed BdMEC to validate its effectiveness in minimizing the execution time of a given task or application. 
\section{Performance Evaluation} 
In this section, the prototype setup and results are discussed, focusing on the performance comparison between the proposed BdMEC and the 
Honeybee model introduced in \cite{honeybee}. The evaluation is conducted based on key parameters which are (a) speed gain, (b) the impact of malicious activities on speed gain, (c) the impact of different applications on speed gain, and (d) privacy preservation of workers.

\subsection{Experimental and Simulation Setup}
In our prototype, we implemented the mobile connectivity and blockchain layers of Fig. \ref{system_model}. The implementation of the cloud layer has been left for future work. The prototype used a CPU with an $11^{th}$ Gen Intel(R) Core(TM) i7-1165G7 @ 2.80GHz, 16.0 GB of installed memory, and Windows 11 Home, as well as four Google Pixel 6 Android devices and one Galaxy A21. Each Google Pixel comes with Internal Memory of 128GB, 12GB of RAM, and Android 12. A21 device comes with Internal Memory of 32GB, 6GB of RAM, and Android 12. A Google Pixel device served as the delegator, while the other devices formed the worker pool. The face match application from \cite{honeybee} was used, which requires searching for human faces in a set of images. One thousand images from flicker \cite{flickerimages} were used with sizes 1-2 MB for which the execution time exceeded five minutes even on a Google Pixel, making it necessary to divide the task into smaller chunks and distribute it among the worker pool for resource sharing. A chunk size of 40 images was used, and the delegator managed the jobs and associated statistics within the system. Wi-Fi Direct was employed as the connection medium to facilitate communication between devices, with the delegator recognizing and communicating with each worker.

\begin{figure}[tp]
\centerline{\includegraphics[width=\linewidth]{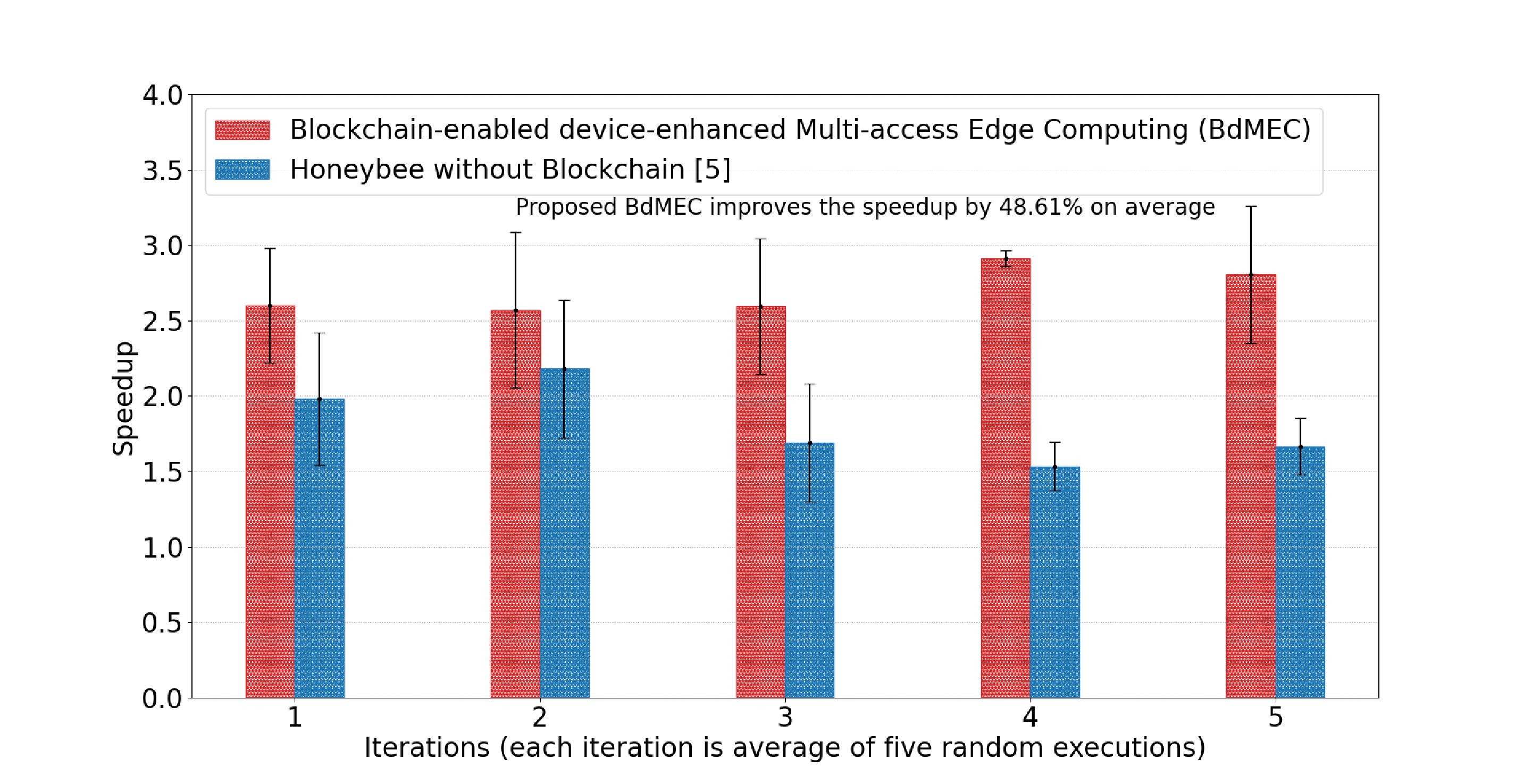}}
\caption{Comparison of speed gain between BdMEC and Honeybee without blockchain \cite{honeybee} over five random executions.}
\label{result1}
\end{figure}

The proposed scenario uses Hyperledger fabric \cite{hyperledger} on a local machine with two nodes to create a realistic blockchain network. Each peer node has a single smart contract (chaincode) installed with one channel called \textit{mychannel}. For simplicity, we implemented the delegator’s ledger twice with one channel to create two ledgers, i.e., the delegator's ledger and the worker's ledger. The SDK version 1.4.22 is used with Caliper 0.4.0 \cite{caliper} to test the blockchain network. The delegator writes the statistics, including the number of jobs executed, the worker's ID, location, speed gain, and behavior $\lambda$, to the blockchain network as a transaction after the task's successful completion. The delegator queries these statistics before selecting a worker from the worker pool $\omega$. The collected statistics are uploaded to the blockchain network in an offline manner. In our prototype implementation, the delegator writes the collected statistics to a text file, which is then used as transaction contents for the blockchain network.

\begin{figure}[!tp]
\centerline{\includegraphics[width=\linewidth]{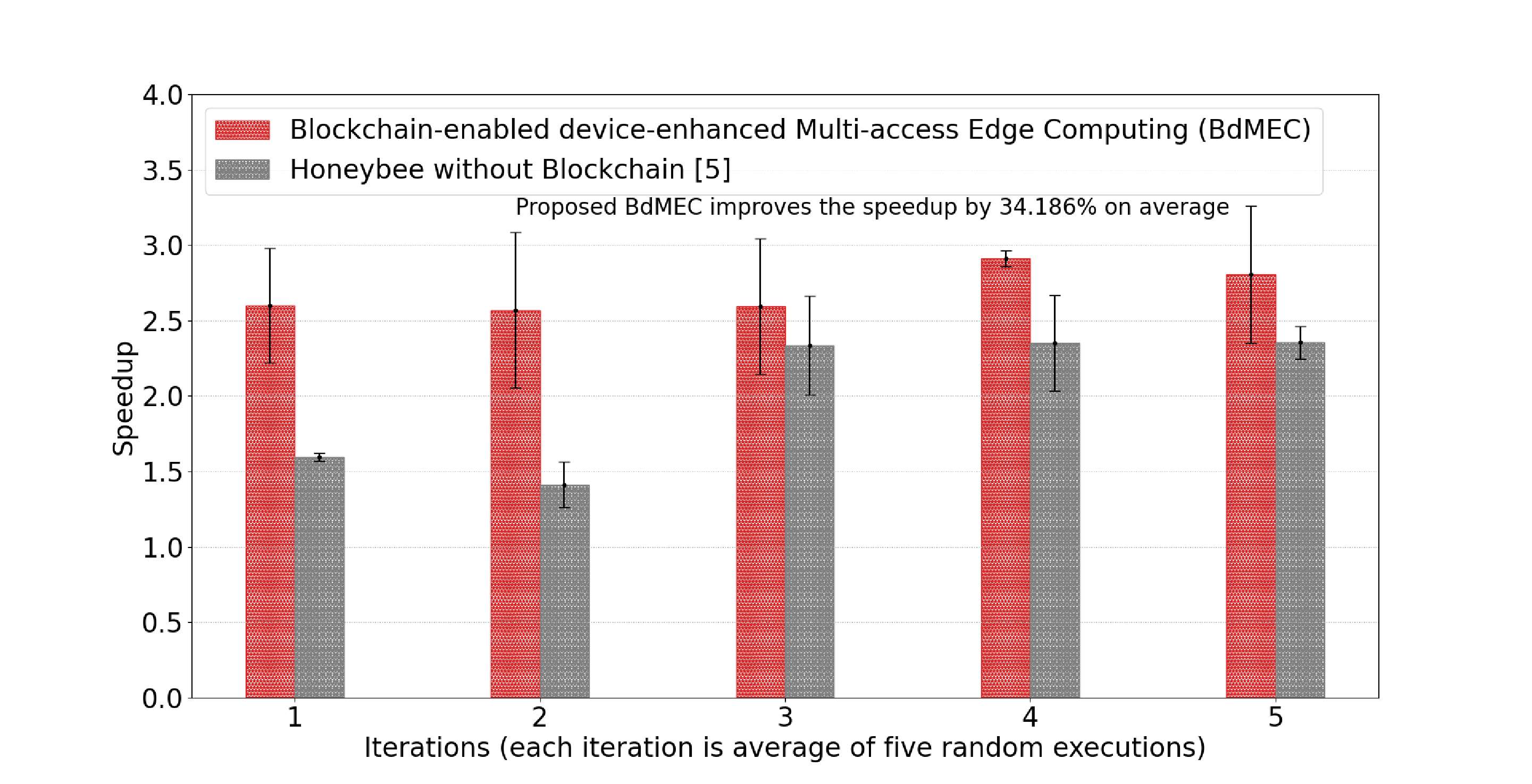}}
\caption{Comparison of speed gain in malicious settings between BdMEC and Honeybee without blockchain\cite{honeybee}.}
\label{result2}
\end{figure}

\begin{figure}[!tp]
\centerline{\includegraphics[width=\linewidth]{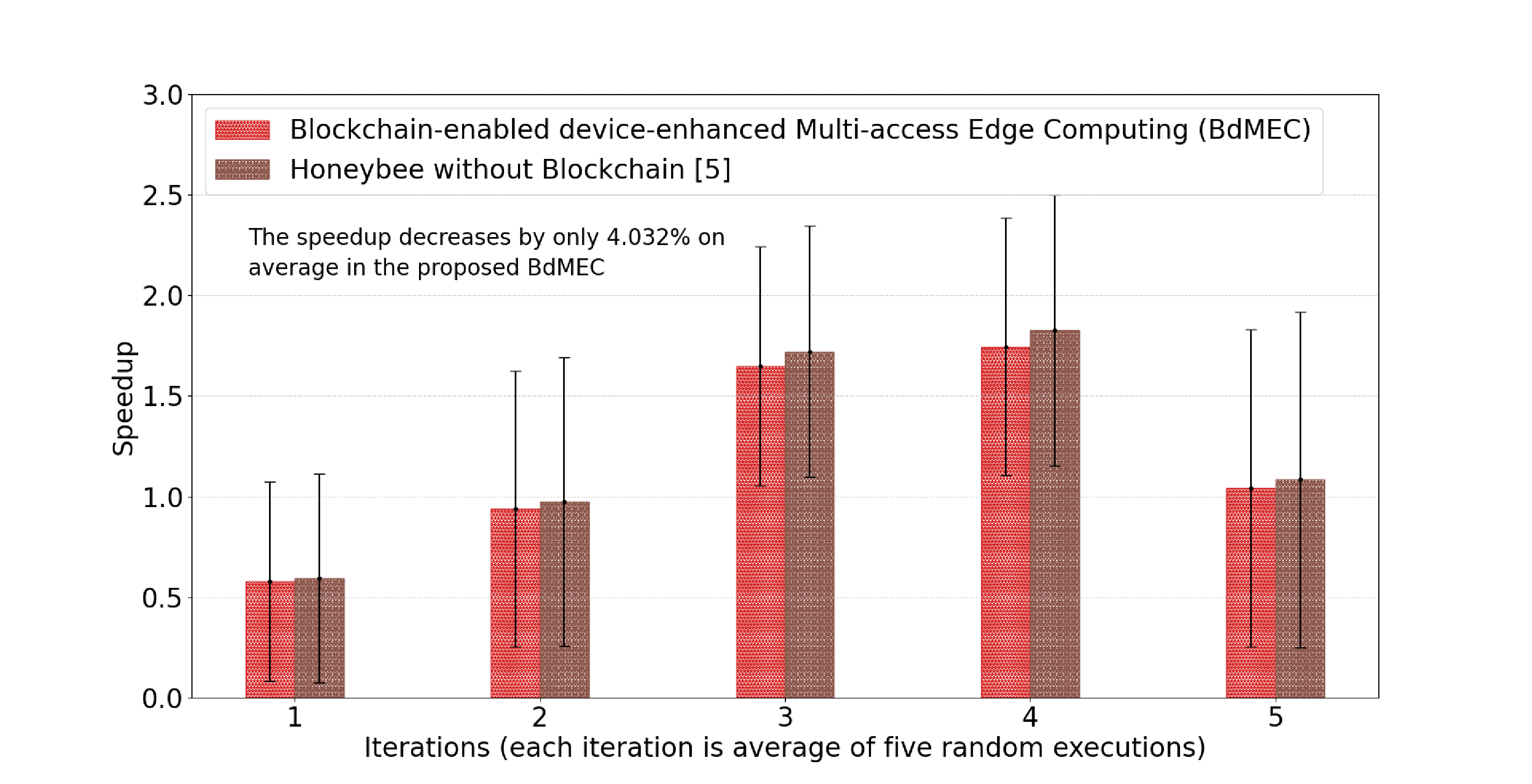}}
\caption{Comparison of speed gain between BdMEC and Honeybee without blockchain \cite{honeybee} over different applications.}
\label{result3}
\end{figure}

\subsection{Results and Discussion}
\paragraph{Speed gain}
The speed gain $\bold{S}$ for both Honeybee without blockchain~\cite{honeybee} and with BdMEC is evaluated and the results are presented in Fig. \ref{result1}. Furthermore, each iteration on the X-axis, i.e., 1-5 is the average value of five random executions with 95\% of confidence interval to smoothen the value of speed gain $\bold{S}$. It is evident from the result that the speed gain $\bold{S}$ for BdMEC is higher than the Honeybee model of \cite{honeybee} for all five iterations. The reason is that BdMEC utilizes the records on blockchain to select workers with relatively high processing capacity based on historical performance, i.e., a faster worker node whereas, in Honeybee \cite{honeybee}, the delegator does not apply a selection criteria for workers and accepts all worker connections. Furthermore, for all five iterations, BdMEC improves the speed gain by 48.61\% on average, hence validating the superiority of BdMEC.  

\begin{table*}[!ht]
\caption{Privacy and average relative error trade-off, and impact of privacy budget $\epsilon$ on precision of selection}
\begin{center}
\begin{tabularx}{\textwidth} { 
  | >{\centering\arraybackslash}X 
  | >{\centering\arraybackslash}X
  | >{\centering\arraybackslash}X
  | >{\centering\arraybackslash}X
  | >{\centering\arraybackslash}X
  | >{\centering\arraybackslash}X | }
\hline
\textbf{Privacy budget $\epsilon$} & \textbf{Relative error \textbf{R} (\%) for worker 1} & \textbf{Relative error \textbf{R} (\%) for worker 2} & \textbf{Precision \textbf{P} (\%) for worker 1} & \textbf{Precision \textbf{P} (\%) for worker 2} \\
\hline
0.01 & 36\% & 48.03\% & 58\% & 88\% \\
\hline
0.1 & 3.16\% & 4.6\% & 92\% & 98\% \\
\hline
0.5 & 0.57\% & 0.92\% & 98\% & 100\% \\
\hline
1 & 0.42\% & 0.38\% & 96\% & 100\% \\
\hline
2 & 0.199\% & 0.206\% & 98\% & 100\% \\
\hline
\end{tabularx}
\label{tab:priv-rel-trad}
\end{center}
\end{table*}

\paragraph{Impact of malicious activities on speed gain}
To expedite time-consuming tasks, workers assist the delegator, but the presence of malicious nodes can complicate the process. These nodes intentionally delay sharing their output, causing tasks to take longer than necessary. We demonstrated this behavior by selecting a Google Pixel device as a malicious node and programming a 50,000-millisecond delay each time it was required to send its result to the delegator. This intentional delay resulted in a prolonged wait for the result and an accumulation of uncompleted tasks in the delegator’s job queue, which were initially assigned to the malicious node. We conducted a series of experiments, and the results are presented in Fig. \ref{result2}. Five iterations were performed, and each iteration represented an average of five random executions. Results shows that BdMEC consistently performed well, demonstrating an average speed gain of 34.186\%. This improvement is achieved by avoiding the assignment of jobs to malicious nodes based on their behavior, which is determined by querying their records stored on the blockchain. In contrast, the Honeybee model ~\cite{honeybee} assigns jobs to malicious nodes, resulting in a negative impact on the overall speed gain. Therefore, our results demonstrate that BdMEC outperforms the Honeybee model.

\paragraph{Impact of different applications on speed gain}
We utilized a different set of four thousand black and white images from \cite{BandWimages} instead of one thousand flicker images in the face match application. The image sizes ranged between 10KB-700KB, which are much smaller than the flicker images. Our objective was to examine the impact of varying job data sizes on the speed gain $\bold{S}$ with BdMEC. The prototype setup was kept the same, and the experiment was conducted with black and white images. The outcome of the experiment is illustrated in Fig. \ref{result3}, with a 95\% confidence interval. The findings indicate that the speed gain $\bold{S}$ for BdMEC is slightly lower than Honeybee in all five iterations. Specifically, the speed gain is decreased by an average of 4.02\% in BdMEC. The reason for this is that the delegator in BdMEC spends more time sending jobs to the workers than the workers save in computation time. Furthermore, given that the image sizes are relatively small (KBs), the results suggest that it is more efficient for the delegator to complete the task themselves rather than offloading it to the workers. 

In addition, the BdMEC model requires additional delay to query the blockchain record for information about a specific worker node before job assignment, which sets it apart from the Honeybee model. Moreover, the average speed gain in this case is lower than the previous two results shown in Fig. \ref{result1} and \ref{result2} for both BdMEC and Honeybee. Specifically, the average speed gain drops from around 2.5 in Fig. \ref{result1} and \ref{result2} to around 1.5 in Fig. \ref{result3}. These results indicate  a diminishing return on offloading when latency is low relative to computation time due to smaller job data (image) size, further compounded by the performance overhead introduced by blockchain integration, rendering the offloading less advantageous.

\paragraph{Privacy preservation of workers}
Differential privacy records noisy counts on the worker's ledger and actual counts on the delegator's ledgers. The privacy budget $\epsilon$ is chosen to mislead malicious and greedy workers, accurately select top workers by the delegator, and enable all workers to track their history in BdMEC, making it transparent.

In our experiments, we used five different values of $\epsilon$ i.e., 0.01, 0.1, 0.5, 1, and 2, and evaluated relative error \textbf{R} (equation \ref{eq:er}) and precision \textbf{P} (equation \ref{eq:prec}) for two workers shown in Table \ref{tab:priv-rel-trad}. The results show that a high value of $\epsilon$ results in small value of average relative error \textbf{R}, i.e., $\epsilon = \{0.5, 1, 2\}$ result in $\textbf{R} = \{0.57\%, 0.42\%, 0.199\%\}$ for worker 1. However, a high value of $\epsilon$ does not give a strong privacy protection guarantee for top workers in the list. On the other hand, results also show that $\epsilon = \{0.5, 1, 2\}$ give a high precision \textbf{P} = $\{0.92, 0.98, 0.96\}$ for the selection of top workers in the list, i.e., 98\% for $\epsilon = 0.5$ for worker 1. 

In addition, it is also evident that with $\epsilon = 0.1$ both worker 1 \& 2 can track their performances with 92\% and 98\% of precision, respectively. Therefore, a suitable value of $\epsilon$ not only protects top workers in the list but also allows workers to track their performance history, thereby making BdMEC transparent to workers.

\section{Conclusion}
We presented a mechanism called Blockchain-enabled device-enhanced MEC (BdMEC) that can deal with diverse devices, even hostile or malicious ones. 
%
Our evaluation of BdMEC using Google pixel devices showed an average speed gain of 48.61\% and 34.186\% by selecting a worker with high capacity and avoiding malicious workers, respectively. 
While our demonstration is with Honeybee, our approach is applicable to   other frameworks.


Future work will expand experiments for a broader array of devices and scenarios, including also energy consumption analysis, and transition from a local machine setup to deploying a real-world blockchain network. 


\bibliographystyle{IEEEtran}
\bibliography{IEEEabrv,blockchain_edgecomp.bib}

\begin{IEEEbiography}[{\includegraphics[width=1in,height=1.25in,keepaspectratio]{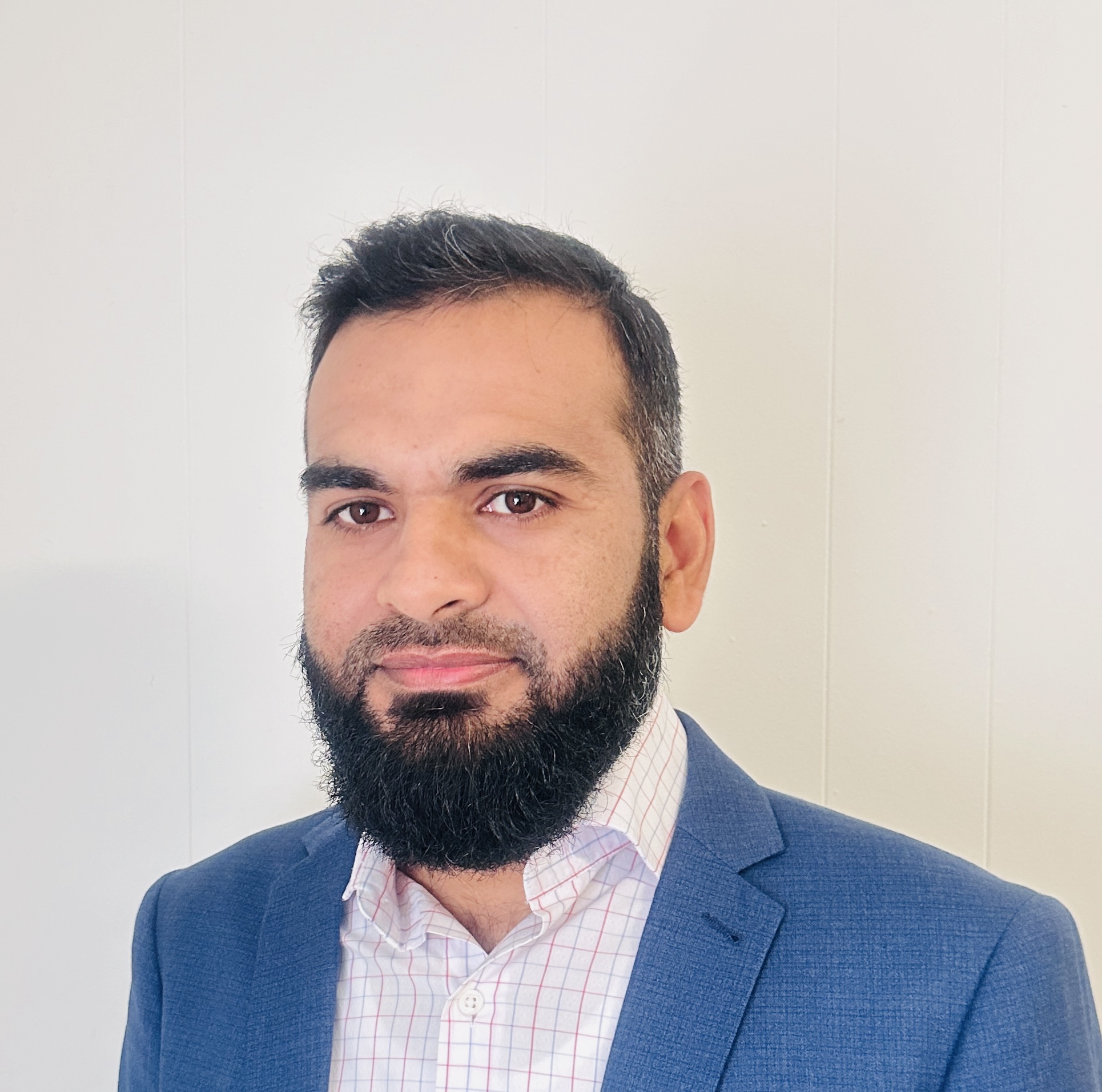}}]{Muhammad Islam, PhD}
is a Research Fellow and Sessional Lecturer in the School of Information Technology at Deakin University, Australia. He has completed his PhD in Computer Science from Swinburne University of Technology, Australia in 2022. His main research areas include Privacy preservation, Differential privacy, Blockchain, and Edge Computing.  
\end{IEEEbiography}
\vspace{-33pt}
\begin{IEEEbiography}[{\includegraphics[width=1in,height=1.25in,keepaspectratio]{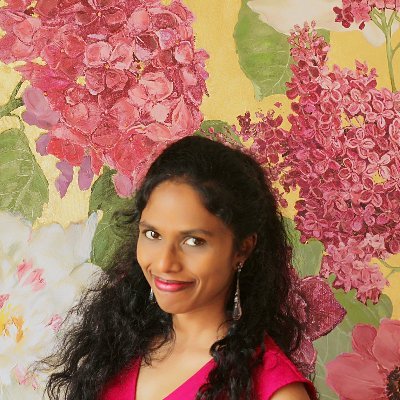}}]{Dr. Niroshinie Fernando}
is a Senior Lecturer in the School of Information Technology at Deakin University, Australia. She completed her PhD in Computer Science from La Trobe University, Australia, in 2015. Her research is centred around the design and development of IoT middleware and systems following software engineering principles to enable long-term adoption, scalability and robustness.    
\end{IEEEbiography}
\vspace{-33pt}
\begin{IEEEbiography}[{\includegraphics[width=1in,height=1.25in,keepaspectratio]{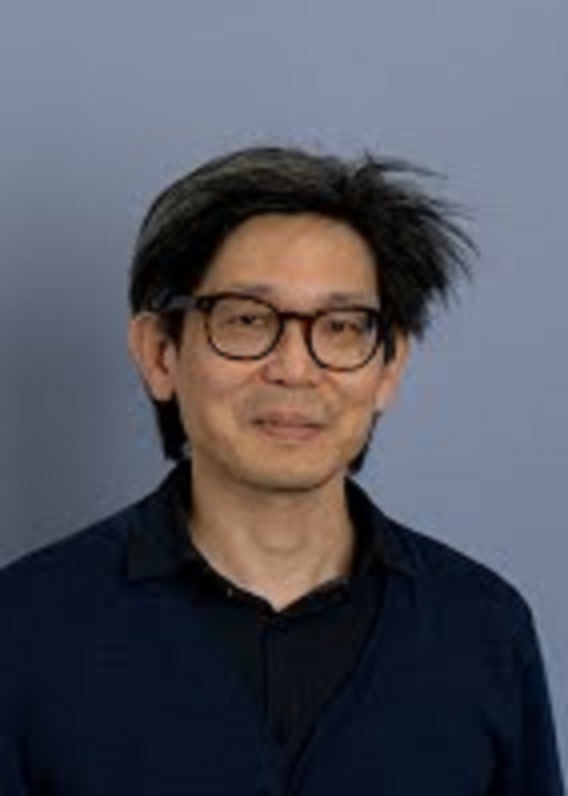}}]{Seng W. Loke} 
is Professor in Computer Science within the School of IT at Deakin University, Australia. He co-directs the IoT Platforms and Applications Lab at Deakin's School of Information Technology  (SIT), and directs the Centre for Software, Systems and Society also in SIT. He received the B.Sc. (First Class Hons.) degree in Computer Science from the Australian National University and the Ph.D. degree in Computer Science from the University of Melbourne, in 1994 and 1998, respectively. His research interests include the Internet of Things, cooperative vehicles, mobile computing, smart city, and quantum Internet computing.    
\end{IEEEbiography}
\vspace{-33pt}
\begin{IEEEbiography}[{\includegraphics[width=1in,height=1.25in,keepaspectratio]{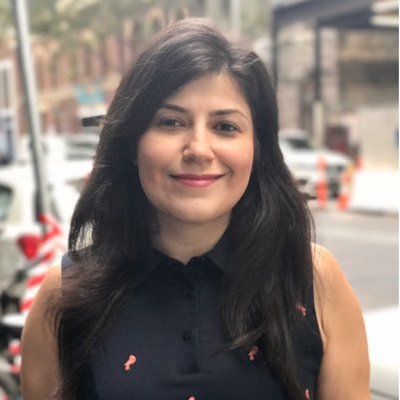}}]{Dr. Azadeh Ghari Neiat}
is a Senior Lecturer in the School of Information Technology at Deakin University, Australia. Prior to her appointment at Deakin in 2019, she was a postdoctoral research fellow and casual lecturer in the School of Computer Science at the University of Sydney. She was awarded her PhD in computer science from RMIT University, Australia, in 2017. Her research interests lie at the intersection of the IoT, Mobile Computing, Crowdsourcing, and Spatio-Temporal Data Analysis.    
\end{IEEEbiography}
\vspace{-18cm}
\begin{IEEEbiography}[{\includegraphics[width=1in,height=1.25in,keepaspectratio]{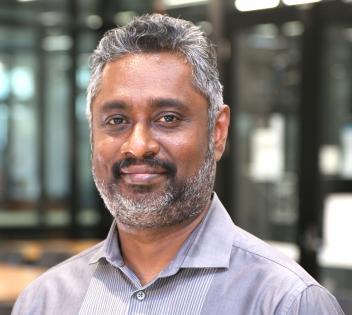}}]{Pubudu N. Pathirana}
(Senior Member IEEE, FEAust) He was a Research Fellow at the Oxford University, U.K.; a Research Fellow at the University of New South Wales, Sydney, Australia; and a Consultant with the Defence Science and Technology Organization, Australia; Visiting Professor with Yale University, USA. He is currently a Full Professor, the Head of Discipline, Mechatronics, Electrical, and Electronic Engineering, and the Director of the Networked Sensing and Biomedical Engineering Research Group, at Deakin University, Australia.    
\end{IEEEbiography}
\end{document}